\documentclass[letterpaper,spanish]{IEEEtran}
\usepackage[T1]{fontenc}
\usepackage[latin9]{inputenc}
\usepackage{color}
\usepackage{babel}
\addto\shorthandsspanish{\spanishdeactivate{~<>}}

\usepackage{array}
\usepackage{textcomp}
\usepackage{amsmath}
\usepackage{graphicx}
\usepackage[unicode=true,
 bookmarks=true,bookmarksnumbered=false,bookmarksopen=false,
 breaklinks=true,pdfborder={0 0 0},pdfborderstyle={},backref=false,colorlinks=true]
 {hyperref}
\hypersetup{pdftitle={Estimación de Poblaciones},
 pdfauthor={William Mendoza y Eduardo NAVAS},
 pdfsubject={Estimación de Poblaciones de Centros de Atención Infantil},
 pdfkeywords={población estadística regresión lineal}}

\makeatletter

\pdfpageheight\paperheight
\pdfpagewidth\paperwidth

\providecommand{\tabularnewline}{\\}

\newenvironment{lyxlist}[1]
	{\begin{list}{}
		{\settowidth{\labelwidth}{#1}
		 \setlength{\leftmargin}{\labelwidth}
		 \addtolength{\leftmargin}{\labelsep}
		 }}
	{\end{list}}

\makeatother

\begin{document}
\title{Estimación de población de centros de atención infantil por medio
de regresiones lineales\thanks{Este artículo fue presentado en el \textbf{Congreso de Electrónca
e Informática 2010} de la \emph{Universidad Centroamericana ``José
Simeón Cañas''}, celebrado el 25 y 26 de noviembre de 2010.}}
\author{William Mendoza, Jefe del Departamento de Matemática, wmendoza@ing.uca.edu.sv\\
Eduardo Adam Navas-López, Catedrático del Departamento de Electrónica
e Informática, enavas@ing.uca.edu.sv\\
Universidad Centroamericana ``José Simeón Cañas''}

\maketitle

\begin{abstract}
Este artículo surge como una alternativa de solución al problema de
estimar la población futura de algunos centros de atención de jóvenes
y niños en riesgo social. La población de estos centros fluctúa mes
a mes debido a diversos factores sociales, económicos y políticos,
por lo que no hay forma de calcular exactamente cuántos jóvenes ingresarán
y cuántos egresarán. Para realizar la estimación, se desarrolló un
modelo matemático para proyectar la población de los centros en el
siguiente mes basándose en regresiones lineales y se implementó en
hoja de cálculo para facilitar su uso.
\end{abstract}


\section{Introducción}

\IEEEPARstart{L}{as} características relevantes de la población de
los centros de atención infantil, según se indagó en la institución
encargada de la administración de estos centros son:
\begin{enumerate}
\item Hay datos de población mensual atendida durante los últimos cinco
años, a partir del año 2005.
\item A veces suceden eventualidades de orden político, social o económico
que hacen variar la población de manera anómala durante normalmente
un mes más o menos: Sucesos como redadas masivas, rescate de grupos
de niños y niñas víctimas de trata de personas, etc..
\item Los eventuales cambios en la legislación hacen que las tendencias
cambien completamente.
\end{enumerate}
Estas poblaciones no responden a modelos de población general como
los modelos exponencial o logísticos, ya que tanto el aumento de la
población como su disminución dependen de condiciones sociales, económicas
y políticas variables y fluctuantes. El crecimiento de estas poblaciones
no está relacionado con el tamaño actual de la población ---característica
básica del crecimiento exponencial\cite{regresi=0000F3n-log=0000EDstica}---
y su decrecimiento no es debido a muertes naturales, por enfermedades
o por agotamiento del ecosistema ---característica del modelo logístico\cite{regresi=0000F3n-log=0000EDstica}---.

Por otro lado, no aplican los métodos de pronóstico como las medias
móviles, ya que estos se basan en que el fenómeno es estacionario\cite{operaciones}
(y los datos de población registrados no muestran patrones repetitivos);
y finalmente las técnicas de suavización exponencial no siempre serían
una buena alternativa ya que estas siempre le dan más ponderación
a los datos más recientes\cite{operaciones}.

Esto hace difícil poder tener una idea de cómo fluctuará la población
en los siguientes meses; sin embargo, eso es fundamental para poder
calcular el presupuesto de alimentación (y de otros rubros) de la
institución encargada de los niños y las niñas.

A pesar de esta dificultad, es altamente prioritario tener una base
objetiva para poder hacer un estimado del presupuesto de alimentación
(y de otros rubros). Por ello, es necesario diseñar y aplicar un modelo
matemático que permita aproximar la población de los centros, y a
partir de esta estimación, calcular el presupuesto de alimentación,
educación, vestuario, etc.

Así, este artículo expone un modelo de estimación de población que
se espera contribuya a objetivizar el cálculo de la población siguiente,
mes a mes, de los centros de atención de jóvenes en riezgo y así ayudar
a calcular más apropiadamente el presupuesto para darles mejor atención.

\section{Metodología}

A continuación se describe el modelo diseñado para responder a las
condiciones particulares del problema de predecir la población de
los centros mencionados.

\subsection{Descripción del Modelo}

\begin{figure*}
\begin{centering}
\includegraphics[scale=0.6]{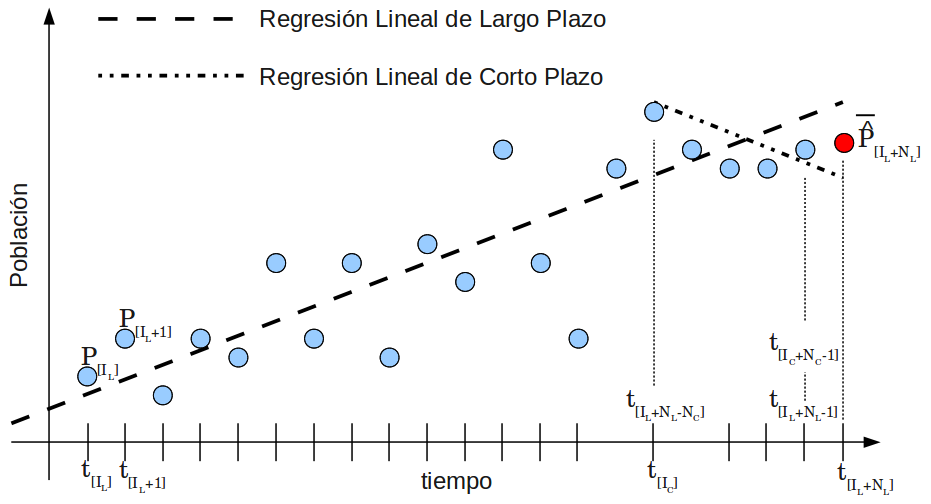}
\par\end{centering}
\caption{Diagrama de las rectas de regresión de largo y corto plazo\label{fig:explicacion-inicial}}

\rule[0.5ex]{1\linewidth}{1pt}
\end{figure*}

El modelo elegido consiste en dos rectas de regresión que se complementan.
Una que considere datos de largo plazo disponibles, que aporte la
información de la tendencia general de la población en el tiempo.
Y otra que considere datos de corto plazo disponibles, que aporte
información de los acontecimientos más recientes que modifican la
tendencia general. Este planteamiento se ilustra en la figura \ref{fig:explicacion-inicial}.

En dicha figura, los círculos azules representan los datos de población
registrados en cada período de tiempo (meses, años, etc.) y el círculo
rojo representa la estimación puntual de la población para el período
de tiempo que se desea estimar.

Si suponemos que los datos de población se irán registrando de manera
consecutiva y sin borrar los datos antiguos, hay que definir desde
cuándo se tomarán datos para la estimación de largo plazo (que en
la figura \ref{fig:explicacion-inicial} es el período de tiempo etiquetado
como $t_{[I_{L}]}$). También hay que definir a partir de cuándo se
considerán datos para la estimación de corto plazo (que en la figura
\ref{fig:explicacion-inicial} es el período de tiempo etiquetado
como $t_{[I_{C}]}$).

Se asume que las dos rectas de regresión se calcularán hasta el último
dato disponible (que en la figura \ref{fig:explicacion-inicial} es
el período de tiempo etiquetado como $t_{[I_{L}+N_{L}-1]}$).

A continuación se presenta la nomenclatura utilizada en el modelo:
\begin{lyxlist}{00.00.0000}
\item [{$N_{L}$:}] Número de períodos de largo plazo a considerar.
\item [{$I_{L}$:}] Índice del período inicial a considerar a largo plazo.
\item [{$N_{C}$:}] Número de períodos de corto plazo a considerar.
\item [{$I_{C}$:}] Índice del primer período de corto plazo a considerar.\\
Asumiendo que el último período será el mismo en la estimación de
corto y largo plazo, se tiene que $I_{L}+N_{L}-1=I_{C}+N_{C}-1$ (véase
la figura \ref{fig:explicacion-inicial}) de lo que se concluye que:
\begin{equation}
I_{C}=I_{L}+N_{L}-N_{C}\label{eq:IC}
\end{equation}
\item [{$P_{[i]}$:}] El valor de la Población en el período $i$-ésimo.\\
\item [{$\overline{\hat{P}}_{\left[I_{L}+N_{L}\right]}$:}] Valor puntual
estimado de la población en el período $I_{L}+N_{L}$ (ver figura
\ref{fig:explicacion-inicial}), considerando la estimación de corto
plazo y la de largo plazo. Nótese que $I_{L}+N_{L}=I_{C}+N_{C}$.
\end{lyxlist}
Adicionalmente, se requiere poder dar mayor peso a la estimación de
corto plazo si los últimos períodos reflejan una tendencia fuerte
que se espera perdure para el período a estimar, y a la inversa, poder
dar mayor peso a la estimación de largo plazo si alguno de los últimos
períodos reflejan una tendencia anómala que no se espera que perdure
para el período a estimar. Para ello, se introduce el siguiente factor:
\begin{lyxlist}{00.00.0000}
\item [{$\alpha_{C}$:}] Peso ponderado para la estimación de corto plazo.
\end{lyxlist}

\subsection{Estimación puntual}

La forma elegida de estimar $\overline{\hat{P}}_{\left[I_{L}+N_{L}\right]}$
es utilizar un promedio ponderado de las estimaciones obtenidas con
las dos regresiones lineales, la de corto plazo y la de largo plazo.
Así: 
\begin{eqnarray}
\overline{\hat{P}}_{\left[I_{L}+N_{L}\right]} & = & \left(A_{L}+B_{L}\cdot t_{\left[I_{L}+N_{L}\right]}\right)\left(1-\alpha_{C}\right)+\nonumber \\
 &  & \left(A_{C}+B_{C}\cdot t_{\left[I_{L}+N_{L}\right]}\right)\alpha_{C}\label{eq:promedio-ponderado}
\end{eqnarray}

Donde los parámetros $A_{L}$ y $B_{L}$ son los parámetros de la
recta de regresión de largo plazo\footnote{Estos parámetros están basados en la forma pendiente-intersepto de
la recta: $Y=A+Bx$}, y los parámetros $A_{C}$ y $B_{C}$ son los parámetros de la recta
de regresión de corto plazo.

Dichos parámetros de regresión se calculan siguiendo las reglas de
regresión lineal, tal como se explica en \cite{Salguero}. Las ecuaciones
resultantes, para calcular los parámetros mencionados son \eqref{eq:B_L}--\eqref{eq:A_C}.

\begin{figure*}
Parámetros de la recta de regresión de Largo Plazo:

\begin{eqnarray}
B_{L} & = & \frac{N_{L}{\displaystyle \sum_{j=0}^{N_{L}-1}t_{\left[I_{L}+j\right]}\cdot P_{\left[I_{L}+j\right]}}-{\displaystyle \left(\sum_{j=0}^{N_{L}-1}t_{\left[I_{L}+j\right]}\right)\left(\sum_{j=0}^{N_{L}-1}P_{\left[I_{L}+j\right]}\right)}}{{\displaystyle N_{L}\sum_{j=0}^{N_{L}-1}\left(t_{\left[I_{L}+j\right]}\right)^{2}-\left(\sum_{j=0}^{N_{L}-1}t_{\left[I_{L}+j\right]}\right)^{2}}}\label{eq:B_L}\\
A_{L} & = & \frac{{\displaystyle \sum_{j=0}^{N_{L}-1}P_{\left[I_{L}+j\right]}}}{N_{L}}-B_{L}\frac{{\displaystyle \sum_{j=0}^{N_{L}-1}t_{\left[I_{L}+j\right]}}}{N_{L}}\label{eq:A_L}
\end{eqnarray}
\end{figure*}

\begin{figure*}
Parámetros de la recta de regresión de Corto Plazo:

\begin{eqnarray}
B_{C} & = & \frac{N_{C}{\displaystyle \sum_{j=0}^{N_{C}-1}t_{\left[I_{C}+j\right]}\cdot P_{\left[I_{C}+j\right]}}-{\displaystyle \left(\sum_{j=0}^{N_{C}-1}t_{\left[I_{C}+j\right]}\right)\left(\sum_{j=0}^{N_{C}-1}P_{\left[I_{C}+j\right]}\right)}}{{\displaystyle N_{C}\sum_{j=0}^{N_{C}-1}\left(t_{\left[I_{C}+j\right]}\right)^{2}-\left(\sum_{j=0}^{N_{C}-1}t_{\left[I_{C}+j\right]}\right)^{2}}}\label{eq:B_C}\\
A_{C} & = & \frac{{\displaystyle \sum_{j=0}^{N_{C}-1}P_{\left[I_{C}+j\right]}}}{N_{C}}-B_{C}\frac{{\displaystyle \sum_{j=0}^{N_{C}-1}t_{\left[I_{C}+j\right]}}}{N_{C}}\label{eq:A_C}
\end{eqnarray}

\rule[0.5ex]{1\linewidth}{1pt}
\end{figure*}

\subsection{Precisión de la estimación}

La estimación puntual en sí misma no es muy útil, puesto que la probabilidad
de que la predicción coincida con el valor real, es casi cero. Para
ello, es necesario definir un \emph{intervalo de confianza} alrededor
de la estimación puntual, es decir un valor máximo y uno mínimo entre
los cuales se espera que se encuentre el valor real con una cierta
probabilidad.

Un propósito adicional de definir un intervalo de confianza es que
en lugar de utilizar la estimación puntual para los cálculos del presupuesto,
se utilice el límite superior del intervalo. Mejor que sobre y no
que falte.

De \cite{folleto} sabemos que el intervalo de confianza para la estimación
de $Y_{n+1}$ por medio de una recta de regresión del tipo $Y=A+B\cdot X$,
está dado por \eqref{eq:intervalo-de-confianza-generico}.

\begin{equation}
\hat{Y}_{n+1}\pm t_{n-2,\alpha/2}^{*}\sqrt{\left[1+\frac{1}{n}+\frac{\left(x_{n+1}-\bar{x}\right)^{2}}{{\displaystyle \sum_{i=1}^{n}x_{i}^{2}}-n\bar{x}^{2}}\right]s_{e}^{2}}\label{eq:intervalo-de-confianza-generico}
\end{equation}

En donde $\hat{Y}_{n+1}$ es la estimación puntual de $Y_{n+1}$,
$t_{n-2,\alpha/2}^{*}$ es el valor\footnote{Denotaremos aquí a la distribución $t$\emph{ de Student} como $t^{*}$
para evitar confusión con la variable independiente de período de
tiempo que estamos utilizando ($t$).} de distribución acumulada de probabilidad $t$ \emph{de Student}
con $n-2$ grados de libertad y $\alpha$ porcentaje de error en la
estimación, $n$ es el número de datos incluídos en la estimación,
$\bar{x}$ es ${\displaystyle \frac{{\displaystyle \sum_{i=1}^{n}x_{i}}}{n}}$,
y finalmente el término $s_{e}^{2}$ es el estimador insesgado de
la varianza de los términos de error (véase \cite{folleto} y \cite{Salguero}),
calculado como:

\begin{equation}
s_{e}^{2}=\frac{{\displaystyle \sum_{i=1}^{n}e_{i}^{2}}}{n-2}=\frac{{\displaystyle \sum_{i=1}^{n}\left(Y_{i}-\hat{Y}_{i}\right)^{2}}}{n-2}=\frac{SCE}{n-2}\label{eq:SCE}
\end{equation}

Al numerador de esta expresión se le conoce como la Suma de los Cuadrados
de los Errores (\emph{SCE}), y el denominador se refiere a los grados
de libertad.\cite{folleto}

Puesto que son dos rectas de regresión en nuestro caso, hay dos radios
para cada intervalo de confianza. Basándose en \eqref{eq:intervalo-de-confianza-generico}
y \eqref{eq:SCE} estos radios son \eqref{eq:radio-largo-plazo} y
\eqref{eq:radio-corto-plazo}.

Finalmente, el intervalo de confianza seleccionado, tomando \eqref{eq:promedio-ponderado},
\eqref{eq:radio-largo-plazo} y \eqref{eq:radio-corto-plazo}, es:\\
\begin{equation}
\overline{\hat{P}}_{\left[I_{L}+N_{L}\right]}\pm\left[\rho_{L}\cdot\left(1-\alpha_{C}\right)+\rho_{C}\cdot\alpha_{C}\right]\label{eq:intervalo}
\end{equation}

\begin{figure*}
Radio del intervalo de confianza de Largo Plazo:

\begin{equation}
\rho_{L}=t_{N_{L}-2,\alpha/2}^{*}\sqrt{\left[1+\frac{1}{N_{L}}+\frac{\left(t_{\left[I_{L}+N_{L}\right]}-{\displaystyle \frac{{\displaystyle \sum_{i=0}^{N_{L}-1}t_{\left[I_{L}+i\right]}}}{N_{L}}}\right)^{2}}{{\displaystyle \sum_{i=0}^{N_{L}-1}\left(t_{\left[I_{L}+i\right]}\right)^{2}}-N_{L}\left({\displaystyle \frac{{\displaystyle \sum_{i=0}^{N_{L}-1}t_{\left[I_{L}+i\right]}}}{N_{L}}}\right)^{2}}\right]\cdot\frac{{\displaystyle \sum_{i=0}^{N_{L}-1}\left(P_{\left[I_{L}+i\right]}-\left(A_{L}+B_{L}\cdot t_{\left[I_{L}+i\right]}\right)\right)^{2}}}{N_{L}-2}}\label{eq:radio-largo-plazo}
\end{equation}
\end{figure*}

\begin{figure*}
Radio del intervalo de confianza de Corto Plazo:

\begin{equation}
\rho_{C}=t_{N_{C}-2,\alpha/2}^{*}\sqrt{\left[1+\frac{1}{N_{C}}+\frac{\left(t_{\left[I_{L}+N_{L}\right]}-{\displaystyle \frac{{\displaystyle \sum_{i=0}^{N_{C}-1}t_{\left[I_{C}+i\right]}}}{N_{C}}}\right)^{2}}{{\displaystyle \sum_{i=0}^{N_{C}-1}\left(t_{\left[I_{C}+i\right]}\right)^{2}}-N_{C}\left({\displaystyle \frac{{\displaystyle \sum_{i=0}^{N_{C}-1}t_{\left[I_{C}+i\right]}}}{N_{C}}}\right)^{2}}\right]\cdot\frac{{\displaystyle \sum_{i=0}^{N_{C}-1}\left(P_{\left[I_{C}+i\right]}-\left(A_{C}+B_{C}\cdot t_{\left[I_{C}+i\right]}\right)\right)^{2}}}{N_{C}-2}}\label{eq:radio-corto-plazo}
\end{equation}

\rule[0.5ex]{1\linewidth}{1pt}
\end{figure*}

Es un intervalo centrado en la estimación puntual que se definió en
\eqref{eq:promedio-ponderado} y con un radio promedio\footnote{Es un promedio ponderado}
entre $\rho_{L}$ y $\rho_{C}$ definidos en \eqref{eq:radio-largo-plazo}
y \eqref{eq:radio-corto-plazo}.

\subsection{Implementación del modelo}

El modelo se implementó como una hoja de cálculo dinámica.

Los datos se colócan de la siguiente manera en alguna hoja:

\begin{tabular}{|c|c|c|c|}
\hline 
 & {\footnotesize{}A} & {\footnotesize{}B} & {\footnotesize{}C}\tabularnewline
\hline 
{\footnotesize{}1} & \textbf{t} & \textbf{Período} & \textbf{Población}\tabularnewline
\hline 
{\footnotesize{}2} & 1 & 2005-enero & $P_{[1]}$\tabularnewline
\hline 
$\vdots$ & $\vdots$ & $\vdots$ & $\vdots$\tabularnewline
\hline 
{\footnotesize{}12} & 11 & 2005-noviembre & $P_{[11]}$\tabularnewline
\hline 
{\footnotesize{}13} & 12 & 2005-diciembre & $P_{[12]}$\tabularnewline
\hline 
$\vdots$ & $\vdots$ & $\vdots$ & $\vdots$\tabularnewline
\hline 
\end{tabular}

En otra hoja, se especifican los parámetros de entrada:

\begin{tabular}{|c|>{\centering}p{0.7\columnwidth}|c|}
\hline 
 & {\footnotesize{}A} & {\footnotesize{}B}\tabularnewline
\hline 
{\footnotesize{}$\vdots$} & \multicolumn{1}{c|}{$\vdots$} & $\vdots$\tabularnewline
\hline 
{\footnotesize{}26} & \textbf{Parámetros Modificables} & \textbf{Valor}\tabularnewline
\hline 
{\footnotesize{}27} & Período inicial a considerar a Largo Plazo & $I_{L}$\tabularnewline
\hline 
{\footnotesize{}28} & Número de períodos a considerar a Largo Plazo & $N_{L}$\tabularnewline
\hline 
{\footnotesize{}29} & Número de períodos a considerar a Corto Plazo & $N_{C}$\tabularnewline
\hline 
{\footnotesize{}30} & Peso ponderado para la estimación de Corto Plazo & $\alpha_{C}$\tabularnewline
\hline 
{\footnotesize{}31} & Porcentaje de confianza del estimador & $\alpha$\tabularnewline
\hline 
\end{tabular}

Para realizar las sumas selectivas y los demás cálculos dinámicos,
se requiere de las siguientes funciones de hoja de cálculo (descripción
adaptada de \cite{lista-de-funciones}):
\begin{itemize}
\item \texttt{FECHA(Año, Mes, Día)}\\
Recibe tres números y devuelve un valor de fecha en función de ellos.
\item \texttt{DIRECCIÓN(Fila, Columna, {[}Modo{]}, {[}Tipo{]}, {[}Hoja{]})}\\
Devuelve una referencia de celda en forma de texto.\\
Los parámetros son:

\begin{lyxlist}{00.00.0000}
\item [{Fila:}] Representa el número de fila de la referencia de celda.
\item [{Columna:}] Representa el número de columna de la referencia de
la celda (el número, no la letra).
\item [{Modo:}] Un número que determina el tipo de referencia devuelta:

\begin{lyxlist}{00.00.0000}
\item [{1:}] Absoluta (\texttt{\$A\$1})
\item [{2:}] Fila absoluta; Columna relativa (\texttt{A\$1})
\item [{3:}] Fila relativa; Columna absoluta (\texttt{\$A1})
\item [{4:}] Relativa (\texttt{A1})
\end{lyxlist}
\item [{Tipo:}] Si se define en \texttt{0}, se utiliza la notación \texttt{R1C1}
\cite{notaci=0000F3n-r1c1}. Si falta este parámetro o se define en
otro valor distinto a \texttt{0}, se utiliza la notación \texttt{A1}
(que es la tradicional).
\item [{Hoja:}] Representa el nombre de la hoja en forma de cadena. Si
se omite, se asume que es una refencia a la misma hoja.
\end{lyxlist}
\item \texttt{INDIRECTO(Referencia, {[}Tipo{]})}\\
Devuelve la referencia especificada por una cadena de texto.\\
Los parámetros son:

\begin{lyxlist}{00.00.0000}
\item [{Referencia:}] Representa una referencia a una celda o a un área
(con formato de texto) para la que se devuelve el contenido.
\item [{Tipo:}] Si se define en \texttt{0}, se utiliza la notación \texttt{R1C1}
\cite{notaci=0000F3n-r1c1}. Si falta este parámetro o se define en
otro valor distinto a \texttt{0}, se utiliza la notación \texttt{A1}
(que es la tradicional).
\end{lyxlist}
\item \texttt{DESREF(Referencia, Filas, {[}Columnas{]}, {[}Alto{]}, {[}Ancho{]})}\\
Devuelve el valor de una celda (o un área), desplazada una determinada
cantidad de filas y columnas desde un punto de referencia concreto.\\
Pueden crearse expresiones dinámicas como: \texttt{SUMA(DESREF(\ldots ))}
para calcular sumas de áreas completas.\\
Los parámetros son:

\begin{lyxlist}{00.00.0000}
\item [{Referencia:}] Es la referencia desde la que la función busca una
nueva referencia.
\item [{Filas:}] Es el número de filas en que la referencia se desplaza
hacia arriba (valor negativo) o hacia abajo (valor positivo).
\item [{Columnas:}] Es el número de columnas en que la referencia se desplaza
hacia la izquierda (valor negativo) o la derecha (valor positivo).
\item [{Alto:}] Es el alto vertical del área que comienza en la nueva posición
de referencia, es decir, el número de filas que contendrá.
\item [{Ancho:}] Es el ancho horizontal del área que comienza en la nueva
posición de referencia, es decir, el número de columnas que contendrá.
\end{lyxlist}
\item \texttt{DIST.T.INV(Probabilidad, GradosdeLibertad)}\\
Calcula el inverso de la distribución $t^{*}$ (La distribución $t$
de Student).\\
Los parámetros son:

\begin{lyxlist}{00.00.0000}
\item [{Probabilidad:}] Es la probabilidad asociada con la distribución
$t$ de dos colas. En nuestro caso, para estimar $t_{N_{L}-2,\alpha/2}^{*}$
y $t_{N_{C}-2,\alpha/2}^{*}$, este parámetro es $1-\alpha$.
\item [{GradosdeLibertad:}] Es el número de grados de libertad de la distribución
$t$. En nuestro caso, son $N_{L}-2$ y $N_{C}-2$.
\end{lyxlist}
\item \texttt{NOD()}\\
Esta función devuelve el valor especial \texttt{\#N/A} (valor no disponible)
que no aparece en los gráficos. Este valor es más apropiado que \emph{cero}
cuando no queremos que un valor aparezca en un gráfico.
\end{itemize}

\subsection{Validación del modelo}

Para verificar la validez del modelo, se aplicó a una serie de datos
históricos de poblaciones de 16 centros.

Para cada centro, se tienen datos históricos de la población mensual
atendida desde enero de 2005, denominado período 1, hasta marzo de
2010, denominado período 63, es decir, 63 períodos.

Para cada centro, se aplicó el modelo para predecir la población,
desde el período 11 (noviembre de 2005) hasta el período 63 (marzo
de 2010), es decir, 53 pruebas por cada centro. En total fueron aplicadas
848 pruebas ($53\times16$).

Los parámetros de entrada establecidos en la prueba fueron los siguientes:
$I_{L}=1$, $N_{C}=6$, $\alpha_{C}=0.5$, $\alpha=0.9$ y $N_{L}$
se hizo iterar desde 10 hasta 62 para estimar la población de los
períodos desde el 11 hasta el 63.

El resultado verificado fue que de las 848 ejecuciones, sólo 112 resultaron
fallidas (13.2\%). Que una ejecución resultara fallida significa que
el valor histórico real de la población resultó fuera del intervalo
de confianza calculado en esa ejecución. Lo que en otras palabras
significa que en esos meses la población sufrió drásticos cambios
más allá de la tendencia registrada, con un incremento (o decremento)
muy por encima de lo normal.

\section{Conclusiones}
\begin{enumerate}
\item Las poblaciones de centros de atención infantil, así como los centros
de reinserción social y otros similares son difíciles de predecir,
debido a que su fluctuación no depende de la población en sí, no tienen
``estaciones'' claramente definidas, y además, su ``ecosistema''
es controlado de forma artificial. Por lo que no responde a los modelos
exponencial, logístico o similares.
\item El modelo propuesto tiene una certeza experimental aceptable de 86.8\%
para estimar la población que se espera atender el siguiente mes,
por lo que puede utilizarse para calcular los presupuestos para los
centros.
\end{enumerate}

\end{document}